\documentclass[aip,apl,amsmath,amssymb,reprint]{revtex4-1}
\usepackage{graphicx}
\usepackage{xcolor}
\usepackage{dcolumn}
\usepackage{bm}

\begin{document}

\title{Quantum oscillations with magnetic hysteresis observed in CeTe$_{3}$ thin films}

\author{Mori~Watanabe}
\affiliation{Department of Physics, Graduate School of Science, Osaka University, Toyonaka, Osaka 560-0043, Japan}
\author{Sanghyun~Lee}
\affiliation{Department of Physics, Graduate School of Science, Osaka University, Toyonaka, Osaka 560-0043, Japan}
\author{Takuya~Asano}
\affiliation{Department of Physics, Graduate School of Science, Osaka University, Toyonaka, Osaka 560-0043, Japan}
\author{Takashi~Ibe}
\affiliation{Department of Physics, Graduate School of Science, Osaka University, Toyonaka, Osaka 560-0043, Japan}
\author{Masashi~Tokuda}
\affiliation{Department of Physics, Graduate School of Science, Osaka University, Toyonaka, Osaka 560-0043, Japan}
\author{Hiroki~Taniguchi}
\affiliation{Department of Physics, Graduate School of Science, Osaka University, Toyonaka, Osaka 560-0043, Japan}
\author{Daichi~Ueta}
\affiliation{Okinawa Institute of Science and Technology Graduate University, Okinawa 904-0495, Japan}
\author{Yoshinori~Okada}
\affiliation{Okinawa Institute of Science and Technology Graduate University, Okinawa 904-0495, Japan}
\author{Kensuke~Kobayashi}
\affiliation{Department of Physics, Graduate School of Science, Osaka University, Toyonaka, Osaka 560-0043, Japan}
\affiliation{Institute for Physics of Intelligence and Department of Physics, The University of Tokyo, Bunkyo-ku, Tokyo 113-0033, Japan}
\author{Yasuhiro~Niimi}
 \email{niimi@phys.sci.osaka-u.ac.jp}
\affiliation{Department of Physics, Graduate School of Science, Osaka University, Toyonaka, Osaka 560-0043, Japan}
\affiliation{Center for Spintronics Research Network, Osaka University, Toyonaka, Osaka 560-8531, Japan}

\date{\today}
\begin{abstract}
We have performed magnetotransport measurements in CeTe$_{3}$ thin films 
down to 0.2~K. It is known that CeTe$_{3}$ has 
two magnetic transitions at $T_{\rm N1} \approx 3$~K and $T_{\rm N2} \approx 1$~K.
A clear Shubnikov-de-Haas (SdH) oscillation was observed at 4~K, 
demonstrating the strong two-dimensional nature in this material.
Below $T_{\rm N2}$, the SdH oscillation has two frequencies, 
indicating that the Fermi surface could be slightly modulated 
due to the second magnetic transition. 
We also observed a magnetic hysteresis in the SdH oscillation below $T_{\rm N1}$. 
Especially, there is a unique spike in the magnetoresistance at $B \approx 0.6$~T only 
when the magnetic field is swept from a high enough field (more than 2~T) to zero field.
\end{abstract}

\pacs{}
\maketitle

Researches on layered materials have attracted much attention over the last 
decade~\cite{geim_nature_2013,novoselov_science_2016}. 
This interest was triggered by the discovery of graphene, where 
not only the polarity but also the density of carriers can be controlled by the electric 
field~\cite{novoselov_science_2004,novoselov_nature_2005,zhang_nature_2005}.
A wide range of materials not only limited to 
semiconductors~\cite{novoselov_pnas_2005,kis_nat_nanotech_2011,ye_science_2012} 
, such as insulators~\cite{dean_nat_nanotech_2010,ponomarenko_nat_phys_2011}, 
superconductors~\cite{cao_nanolett_2015,fese_nmat_2015,shiogai_nphys_2016,zhang_nature_2019}, 
and ferromagnetic 
materials~\cite{gong_nature_2017,huang_nature_2017,wang_nnano_2018,fei_nmat_2018,deng_nature_2018}, 
have been actively studied with the aim to control the physical properties 
or the phase transition temperature by applying the electric field to such thin film devices.
More recently, the research field, so-called van der Waals engineering, 
has become an important stream~\cite{geim_nature_2013,novoselov_science_2016}. 
The most striking discovery is the superconductivity in twisted bilayer graphene, 
which is an originally zero gapped semiconductor~\cite{cao_nature_2018}.
By stacking a strong spin-orbit transition metal dichalcogenide on 
a ferromagnetic thin layer, a magnetic skyrmion phase can be induced~\cite{arxiv_2019}. 
Thus, it is an urgent task to investigate a variety of materials which can be fabricated into 
atomically thin films and explore atomically stacked devices with new physical properties.
Especially, magnetic materials could be useful for future spintronic 
applications~\cite{song_science_2018,wang_sci_adv_2018}.

\begin{figure}
\begin{center}
\includegraphics[width=6.5cm]{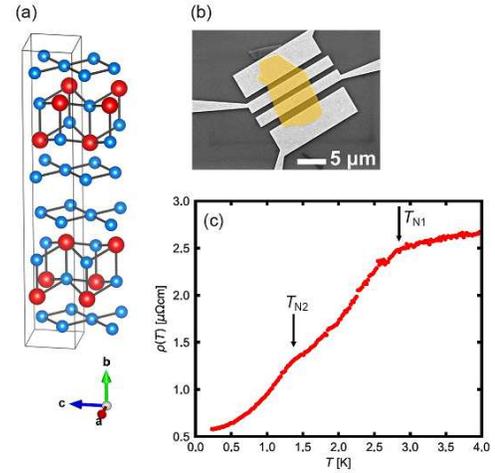}
\caption{(a) Crystal structure of CeTe$_{3}$. 
The red and blue spheres represent Ce and Te atoms, respectively.
The black lined box indicates the unit cell, with unit lattice vectors 
\textit{a} $\sim$ \textit{c} $\sim$ 4.4~\AA~and 
\textit{b} $\sim$ 26~\AA.
(b) SEM image of a typical thin film device. CeTe$_{3}$ is shown in yellow (false color).
(c) Temperature dependence of the resistivity in the CeTe$_{3}$ thin film device. 
There are two resistivity drops at 2.7~K and 1.3~K, 
which correspond to the two magnetic transition temperatures $T_{\rm N1}$ and 
$T_{\rm N2}$ respectively.}
\label{figure1}
\end{center}
\end{figure}

CeTe$_{3}$ is a layered material in the family of rare earth ($R$) tritellurides, i.e., $R$Te$_{3}$. 
It is known as a heavy fermion system with a localized 4$f^{1}$ orbital at the Ce$^{3+}$ site. 
Its crystal structure consists of a NaCl-type CeTe layer which is responsible for 
its magnetic properties, separated by two Te sheets 
which are responsible for the highly two-dimensional (2D) electric transport~\cite{iyeiri_prb_2003}, 
as shown in Fig.~\ref{figure1}(a). 
Due to the highly 2D electrical transport, the material forms an incommensurate 
charge density wave (CDW) from well above room temperature, which has been studied 
extensively~\cite{dimasi_prb_1995,brouet_prl_2004,kim_prl_2006,
brouet_prb_2008,christos_jacs_2006,tomic_prb_2009,ralevic_prb_2016}.

This material is also known to show two magnetic phase transitions 
at low temperatures~\cite{iyeiri_prb_2003,deguchi_jp_2009,zocco_prb_2009,okuma_sci_rep_2020}. 
The first magnetic transition at $T_{\rm N1} = 3.1$~K is understood to 
be from a paramagnetic state to an 
antiferromagnetic (possibly short range ordering) state. 
In this phase, the magnetic moment at the Ce site is antiferromagnetically 
coupled and aligned to an easy axis perpendicular to the layer stacking direction. 
The second magnetic transition is known to be another antiferromagnetic 
(possibly long range ordering) transition~\cite{iyeiri_prb_2003,deguchi_jp_2009} 
at $T_{\rm N2} = 1.3$~K. 
Unlike the case of the first transition, 
a clear peak in the heat capacity has been observed below $T_{\rm N2}$ 
and the magnetic moment is still aligned to the in-plane direction (called \textit{non-parallel} easy axis), 
but different from the easy axis in the first transition~\cite{zocco_prb_2009}. 
Nevertheless, the details of these magnetic ordering states are poorly understood. 
Furthermore, there are no reports on thin film transport measurements.

\begin{figure}
\begin{center}
\includegraphics[width=6.5cm]{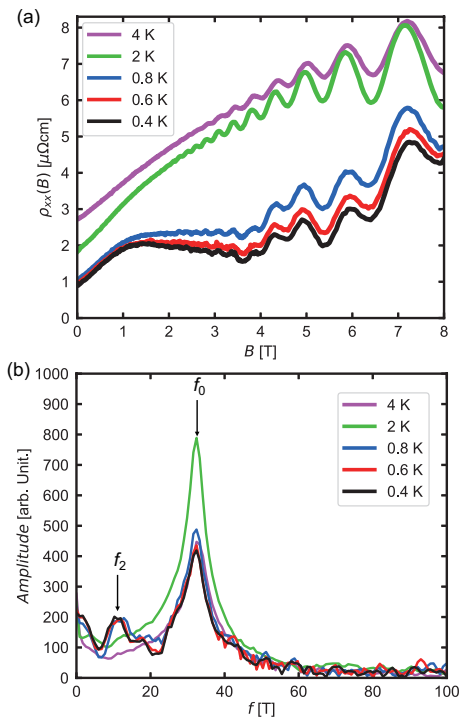}
\caption{(a) Magnetoresistance measured with a 40~nm thick CeTe$_{3}$ thin film device 
at several different temperatures. The external magnetic field was applied perpendicular to the plane 
and swept from zero to 8~T. 
(b) FFT for the derivative of $\rho_{xx}(B)$ versus $1/B$. 
The main oscillation peak has been observed at $f_{0} = 32.4$~T below 4 K. 
The secondary oscillation peak has been observed at $f_{2} = 11.0$~T only below 0.8~K. 
The peak at 0~T is due to the DC offset of the frequency.}
\label{figure2}
\end{center}
\end{figure}

\begin{figure}
\begin{center}
\includegraphics[width=6cm]{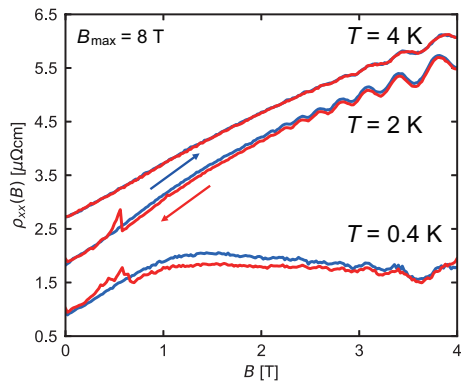}
\caption{Hysteresis in magnetoresistance measured at $T = 2$ and 0.4~K. 
There is a clear hysteresis between the blue and red curves 
where the magnetic field is swept from 0 to 8~T and from 8 to 0~T, respectively. 
A clear jump in resistance was observed only for the red curve at $B \approx 0.6$~T. 
Such a hysteresis was not observed at $T = 4$~K above $T_{\rm N1}$.}
\label{figure3}
\end{center}
\end{figure}

In the present work, we have performed magnetotransport measurements 
with 30-40 nm thick CeTe$_{3}$ devices. We have observed the coexisitance of quantum oscillation with unique magnetotransport phenomena. Specifically,  
a clear Shubnikov-de-Haas (SdH) oscillation was observed 
in the magnetic field $B$ range from 3 to 8~T even above $T_{\rm N1}$, 
demonstrating the existence of a small Fermi surface pocket. 
Since such a SdH oscillation has never been reported in bulk CeTe$_{3}$,
the result reveals its strong 2D nature which is possibly enhanced by the thin film fabrication. Furthermore, 
the SdH oscillation has two frequencies below $T_{\rm N2}$. 
This could originate from the modification of the Fermi surface due to 
the second magnetic transition at $T_{\rm N2}$. 
We also observed a magnetic hysteresis in the SdH oscillation below $T_{\rm N1}$. 
In particular, a sharp resistance peak appears at $B \approx 0.6$~T only when 
the magnetic field is swept from a high enough magnetic field (more than 2~T) 
to zero field (see Fig.~\ref{figure3}). 

Single crystals of CeTe$_{3}$ were synthesized in an evacuated quartz tube. 
The tube was heated up to 900$^\circ$C and slowly cooled to 550$^\circ$C over a period of 7 days. 
To fabricate thin film devices from the bulk CeTe$_{3}$, 
we used the mechanical exfoliation technique using scotch 
tapes~\cite{novoselov_science_2004,novoselov_nature_2005,zhang_nature_2005,novoselov_pnas_2005}.
It is noted that all the following fabrication processes should be carried out inside a glove box 
with Ar purity of 99.9999\%, since CeTe$_{3}$ is extremely sensitive to ambient air.
After the mechanical exfoliation process, many CeTe$_{3}$ thin flakes on the scotch tape 
were transferred onto a thermally-oxidized silicon substrate. 
We then spin-coated polymethyl-methacrylate (PMMA) resist onto the substrate. 
The substrate was taken out from the glove box and electrode patterns were printed 
using electron beam lithography. 
After the lithography, the substrate was put back into the glove box again 
for the development of the resist. 
The Au electrodes were deposited using electron beam deposition in a vacuum chamber 
next to the glove box. 
Before the deposition of Au, Ar milling was performed to remove 
the residual resist and any possibly oxidized layers of CeTe$_{3}$. 
It should be noted that contact resistance varies greatly with electrode material. 
Au electrodes were found to have a minimum contact resistance 
compared to Ti/Au or Cu. 
In order to avoid further damage in ambient condition after the fabrication, 
the device was capped with the PMMA shortly after the electrode deposition and the lift-off process. 
Figure~\ref{figure1}(b) shows a scanning electron microscopy (SEM) image of one of the 
thin film devices. 

Electrical transport measurements were performed by the conventional four-probe method 
using a lock-in amplifier. The device was cooled with a $^3$He/$^4$He dilution refrigerator 
down to 0.2 K and the external magnetic field was applied using a superconducting magnet.
The thicknesses of all measured CeTe$_{3}$ thin films were confirmed 
by using a commercially available atomic force microscope after finishing all the 
electrical measurements.

The resistivity $\rho (T)$ measured with the CeTe$_{3}$ thin film device 
in Fig.~\ref{figure1}(b) is plotted as a function of temperature in Fig.~\ref{figure1}(c). 
Although the CeTe structure is half-metallic on its own, the high conductivity 
between the two Te sheets gives rise to a highly metallic temperature dependence 
in CeTe$_{3}$~\cite{deguchi_jp_2009}. 
There are two resistivity changes in the low temperature region:
the first resistivity drop at $T = 2.7$~K, and the second resistivity kink 
structure at $T = 1.3$~K. 
These behaviors are consistent with the anomalies observed 
in bulk CeTe$_{3}$ resistivity measurements, which correspond to the two magnetic phase transitions 
at $T_{\rm N1} = 3.1$~K and $T_{\rm N2} = 1.3$~K~\cite{iyeiri_prb_2003,deguchi_jp_2009}. 
The resistivity at room temperature of this device is 81.3~$\mu\Omega\cdot$cm, 
resulting in the residual resistivity ratio (RRR) of 59.2 with respect to $\rho (T={\rm 1.5~K})$. 
This is 1.3 times higher than the previously reported value of 44.9~\cite{iyeiri_prb_2003}. 
At the lowest temperature ($T = 0.2$~K), RRR reaches a value of 140, indicating that 
the device is a high quality single crystal CeTe$_{3}$ thin film.

We next performed magnetoresistance measurements up to $B = 8$~T 
for the temperature range from 0.4 to 4~K, as shown in Fig.~\ref{figure2}(a). 
The external magnetic field $B$ was applied perpendicular to the plane, i.e., along 
the $b$-axis of CeTe$_{3}$ and swept from zero to 8~T. 
A large positive magnetoresistance $\rho_{xx}(B)$ was observed in the low field regime ($B < 1$~T) 
along with a clear SdH oscillation which develops from magnetic fields as low as $B = 2$~T 
in the case of $T = 2$~K. 
As far as we know, such a quantum oscillation (including the de Haas-van Alphen effect) 
has never been reported so far even for bulk CeTe$_{3}$, 
although Lei \textit{et al}. have recently reported a SdH oscillation in 
GdTe$_{3}$ thin films~\cite{lei_sci_adv_2020}.
This fact is possibly related to much stronger 2D nature 
in thin films, compared to bulk crystals. 
Furthermore, the magnetoresistance behavior drastically changes 
below $T = 0.8$~K, which is below the second magnetic transition temperature $T_{\rm N2}$. 

In order to extract the oscillatory part of the magnetoresistance, 
the derivative of $\rho_{xx}(B)$ with respect to $1/B$ was obtained numerically. 
Fast Fourier transform (FFT) was then performed in order to obtain the frequency $f$ of 
the quantum oscillation. The results are presented in Fig.~\ref{figure2}(b). 
The main oscillation observed at all temperatures 
corresponds to a frequency of $f_{0} = 32.4$~T, 
which is about two times smaller than that of GdTe$_{3}$~\cite{lei_sci_adv_2020}. 
Below $T = 0.8$~K, there is an additional oscillation with $f_{2} = 11.0$~T, 
which can be seen as the secondary peak in the FFT spectrum in Fig.~\ref{figure2}(b). 
The frequencies of these oscillations are proportional to the extremal cross sectional areas 
of the Fermi surface perpendicular to the applied magnetic field: 
$\displaystyle S = 2\pi e f/\hbar$,
where $S$ is the Fermi surface area, $e$ is the elementary charge, 
$f$ is the frequency in the unit of T, 
and $\hbar$ is the reduced Planck constant~\cite{lifshitz_jetp_1956}. 
Using this equation, we obtained the Fermi surface areas of 
$S_{0} = 3.09 \times 10^{13}$~cm$^{-2}$ for $f_{0} = 32.4$~T, 
and $S_{2} = 1.05 \times 10^{13}$~cm$^{-2}$ for $f_{2} = 11.0$~T. 
The main oscillation 
is present even above the two magnetic transition temperatures,
but well below its CDW transition temperature 
($T_{\rm CDW} >$ 500~K)~\cite{brouet_prl_2004,kim_prl_2006,brouet_prb_2008,christos_jacs_2006}. 
The origin of this Fermi surface pocket can be attributed to 
the reconstruction of the Fermi surface due to the incommensurate CDW, 
which has been observed through photoemission spectroscopy 
experiments~\cite{brouet_prl_2004,brouet_prb_2008} 
as well as through quantum oscillations of other $R$Te$_{3}$ 
materials~\cite{lei_sci_adv_2020,ru_prb_2008,sinchenko_JLTP_2016}. 

Although the angular dependence of the SdH oscillation has not been measured 
in the present study, it was already demonstrated in a similar tritelluride thin film device, 
i.e., GdTe$_{3}$~\cite{lei_sci_adv_2020} where the SdH oscillation follows $1/\cos\theta$ 
($\theta$ is the angle between the applied magnetic field and the layer stacking direction). 
This indicates a highly 2D geometry of the Fermi pockets. 
Since the crystal structure and the high conduction of the Te layers are 
the same for GdTe$_{3}$ and CeTe$_{3}$, 
we believe that the SdH oscillation observed in CeTe$_{3}$ also originates from 
a highly 2D Fermi surface pocket. Especially, $S_{0}$ in the CeTe$_{3}$ device 
is two times smaller than that in a thin film GdTe$_{3}$ device, 
where the effective mass is known to be as small as $\approx 0.1m_{0}$ 
($m_{0}$ is the bare electron mass)~\cite{lei_sci_adv_2020}.
This fact suggests that the conduction electrons between the
Te sheets in the CeTe$_{3}$ device have a similarly small effective mass. 
In addition, we detected the secondary oscillation with a frequency of $f_{2}$, 
which does not exist in GdTe$_{3}$ and develops 
only after the second magnetic transition temperature $T_{\rm N2}$. 
According to Ref.~\onlinecite{deguchi_jp_2009}, 
the second magnetic transition at $T_{\rm N2}$ is related to 
a spin density wave transition with formation of heavy quasiparticles. 
Thus, such reconstruction of the Fermi surface along with this transition 
could be a possible cause of this new oscillation, but
further investigation is required to confirm the hypothesis. 
We note that these SdH frequencies have also been observed in 
multiple different CeTe$_{3}$ devices, confirming its reproducibility.

In a typical SdH oscillation, the FFT amplitude for a given frequency 
decreases with increasing temperature. However, FFT amplitudes of the main oscillation 
remained mostly constant, with the exception at $T = 2$~K.
A similar result has been reported in GdTe$_{3}$, 
where the FFT amplitude plateaus below its antiferromagnetic transition 
temperature, while showing a typical temperature dependence well above 
the transition temperature~\cite{lei_sci_adv_2020}. 
Our scenario is consistent with this report: in other words, 
the unconventional temperature dependence of the 
FFT amplitude depends strongly on the interaction between the magnetic order at the CeTe site 
and the conduction electrons. 

\begin{figure}
\begin{center}
\includegraphics[width=7cm]{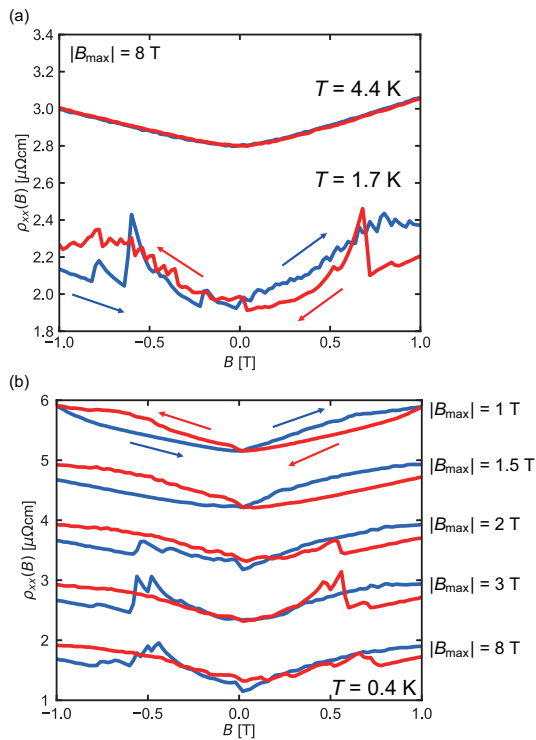}
\caption{ Measurements performed on a different CeTe$_{3}$ device. (a) Magnetoresistance in the field range of $\pm 1$~T measured 
at $T = 4.4$ and 1.7~K.
A similar resistance jump to Fig.~\ref{figure3} was observed both at $B \approx \pm 0.6$~T. 
In this case, $|B_{\rm max}| = 8$~T. 
(b) $|B_{\rm max}|$ dependence of the magnetoresistance in the field range of $\pm 1$~T 
measured at $T = 0.4$~K. The sharp peak disappears when $|B_{\rm max}|$ is smaller than 1.5~T.}
\label{figure4}
\end{center}
\end{figure}

In addition to the SdH oscillations discussed above, 
we observed a magnetic hysteresis behavior, superimposed onto the SdH oscillations, 
below the first antiferromagnetic ordering temperature $T_{\rm N1}$. 
This is highlighted in Fig.~\ref{figure3}. 
At 4~K above $T_{\rm N1}$, there is no hysteresis in the SdH oscillation. 
Below $T_{\rm N1}$, however, a clear magnetic hysteresis is observed with 
a sharp peak structure at $B \approx 0.6$~T. 
The magnetic hysteresis in the SdH oscillation vanishes above $B \approx 4$~T, and 
the sharp peak appears only when $B$ is swept from a high enough field 
(in this case, $B_{\rm max} = 8$~T) to zero field, which we call the negative field sweeping. 
Note that quantum oscillations with magnetic hysteresis are 
reproducible for other CeTe$_{3}$ thin film devices, even below $T_{\rm N2}$, and also 
when the applied magnetic field is negative (see Fig~\ref{figure4}). 
Furthermore, the peak amplitude depends on the absolute value of 
$B_{\rm max}$. At $T = 0.4$~K, when $|B_{\rm max}|$ is smaller than 1.5~T, 
the peak at $B \approx 0.6$~T vanishes, while still preserving the hysteresis behavior, 
as shown in Fig.~\ref{figure4}(b).
Such a hysteresis has never been reported previously 
in bulk CeTe$_{3}$ as well as in thin film GdTe$_{3}$ devices~\cite{lei_sci_adv_2020}.
On the other hand, a hysteresis closely resembling our measurements have been observed 
for the magnetoresistance measurements in CeTe$_{2}$~\cite{jung_prb_2000}, 
a variation of CeTe$_{3}$. CeTe$_{2}$ has the same crystal structure as CeTe$_{3}$ 
but has a single Te layer instead of the double Te layers, 
and is known to order ferrimagnetically with an easy axis along the layer stacking 
direction~\cite{park_pb_1998}. This is different from the magnetic order of 
bulk CeTe$_{3}$, where the magnetic moments at the Ce sites in the two magnetic phases 
are believed to be aligned to the basal plane~\cite{deguchi_jp_2009,zocco_prb_2009}. 
One possible scenario to explain the magnetoresistance hysteresis and peak structure 
observed in CeTe$_{3}$ devices is that by the thin film fabrication, 
we have induced a canting of the magnetic moments along the layer stacking direction, 
resulting in a similar magnetoresistance effect to CeTe$_{2}$. 
There is a supportive result where the perpendicular anisotropy of a van der Waals ferromagnet 
Fe$_{5}$GeTe$_{2}$ is enhanced by making it atomically thinner~\cite{ohta_apex_2020}.
In order to elucidate more details, it would be desirable to perform 
further experiments in future about the thickness dependence of the peak structure. 

In conclusion, we have performed magnetotransport measurements 
of 30-40~nm thick CeTe$_{3}$ thin film devices. 
A clear SdH oscillation was observed from $T = 4$~K, 
indicating a highly two-dimentional character of the conduction electrons, possibly enhanced due to thin film fabrication. 
Below the second magnetic transition temperature $T_{\rm N2}=1.3$~K, on the other hand, 
SdH oscillations with two different frequencies were obtained.
The FFT analysis revealed the existence of two small Fermi pockets 
whose sizes are $3.09 \times 10^{13}$~cm$^{-2}$ and $1.05 \times 10^{13}$~cm$^{-2}$. 
In addition, a magnetic hysteresis superimposed to the SdH oscillation was detected 
below the first antiferromagnetic temperature $T_{\rm N1}=2.7$~K. 
Especially, a sharp peak at $B \approx 0.6$~T was clearly observed 
when the magnetic field was swept from the high enough field to zero field. 
Materials where quantum oscillations and magnetic hysteresis coexist are extremely scarce. 
Along with the ease of thin film fabrication through mechanical 
exfoliation, further researches on CeTe$_{3}$ 
could pave the way for $f$-orbital spintronics and could provide an ideal stage 
for understanding the interplay between electronic quantum conduction 
and localized heavy fermion spins.

We thank H. Sakai, K. Kuroki, M. Ochi, and K. Deguchi for fruitful discussions. 
The cell structure of  CeTe$_{3}$ was visualized using VESTA~\cite{vesta}.
This work was supported by JSPS KAKENHI 
(Grant Numbers JP16H05964, JP17K18756, JP19K21850, JP20H02557, JP26103002, 
JP19H00656, JP19H05826), 
Mazda Foundation, Shimadzu Science Foundation, 
Yazaki Memorial Foundation for Science and Technology, 
SCAT Foundation, Murata Science Foundation, Toyota Riken Scholar, 
and Kato Foundation for Promotion of Science.

The data that support the findings of this study are available from the corresponding author 
upon reasonable request.

\end{document}